\documentclass[nocopyright]{cimento}

\usepackage{amsmath,graphicx}

\title{Two-particle Bose-Einstein correlations and their Lévy parameters in PbPb collisions at 5.02 TeV}
\author{M.~Csan\'ad for the CMS Collaboration}
\instlist{\inst{} ELTE E\"otv\"os Lor\'and University, Department of Atomic Physics, P\'azm\'any P\'eter s\'et\'any 1/a, Budapest, H-1117, Hungary}
\begin{document}

\maketitle

\begin{abstract}
In these proceedings we discuss the measurement of Bose-Einstein momentum correlation function of pairs of charged hadrons in PbPb collisions at $\sqrt{s_{_\mathrm{NN}}}=5.02$ TeV. We describe the measured correlations with correlation functions derived from Lévy type source distributions. Using a transverse momentum and centrality binning, we extract the correlation strength parameter $\lambda$, Lévy index $\alpha$ and Lévy scale parameter $R$ as a function of pair transverse mass $m_\mathrm{T}$, for various centralities.
\end{abstract}

\section{Introduction}
Measuring Bose-Einstein or Fermi-Dirac correlations provides an indispensable tool to understand the femtometer scale geometry of particle (boson or fermion) production in ultrarelativistic nuclear collisions~\cite{PHENIX:2017ino,CMS:2023xyd}. These correlations can be understood on the basis of the HBT effect~\cite{HanburyBrown:1952na} to arise from the wave nature of the observed particles, or the quantum-statistical nature of particle detection events. If one neglects final-state interactions, then for identical bosons, the connection between the observed two-particle momentum correlations ($C(q)$) and spatial particle creation distribution ($S(r)$, referred to as ``source'') is as follows:
\begin{align}
C(q) = 1 + \frac{\tilde S(q)}{\tilde S(0)},
\end{align}
where $q$ is the momentum difference of the pair, and $\tilde S$ is the Fourier transform of $S$ (and in the denominator, it is taken at $q=0$, producing the integral of $S$). In case of a dynamic source (as is the case in ultrarelativistic nuclear collisions), both $S$ and $C$ depend on the average momentum of the pair ($K$). In this case $S$ is often assumed to have a parametric form, and the parameters would then depend on $K$. 

For a source with a L\'evy-stable distribution~\cite{PHENIX:2017ino,Csorgo:2003uv}, one then obtains
\begin{align}
C(q,K) = 1 + \lambda(K)\exp\left(-(qR(K))^{\alpha(K)}\right),
\end{align}
where $R$ is the (momentum-dependent) spatial scale of the source, $\alpha$ is its momentum-dependent L\'evy index, and $\lambda$ is the (also momentum-dependent) strength of the source. Parameter $\lambda$ needs to be defined (and may differ from unity) due to multiple reasons, including particle production through decays~\cite{Csorgo:1994in,PHENIX:2017ino} or lack of particle identification~\cite{CMS:2023xyd}.

The appearance of L\'evy-stable distributions may have its roots in several physical phenomena~\cite{Csanad:2024hva}: (i) anomalous diffusion~\cite{Metzler:1999zz,Csanad:2007fr}, (ii) jet fragmentation~\cite{Csorgo:2004sr}, (iii) a second-order phase transition~\cite{Csorgo:2005it}, (iv) or resonance decays~\cite{Csanad:2007fr,Kincses:2022eqq,Korodi:2022ohn}. In case of ultra-relativistic PbPb collisions, points (i) and (iv) may play a role. See the above references for more details.

In addition, final-state interactions may also play an important role in the momentum correlations of identical particle pairs. In particular, the Coulomb interaction has to be taken into account.~\cite{Nagy:2023zbg} Note that the assumption of L\'evy-stable sources is important also when one investigates the collision geometry, collision energy, or pair momentum dependence of $R$, as with a Gaussian assumption (corresponding to $\alpha=2$) the shape and scale changes may be entangled~\cite{Csanad:2024hva}. Furthermore, it also has been shown in Refs.~\cite{Csanad:2024hva,Kurgyis:2020vbz} that a one-dimensional measurement is adequate for these investigations for an approximately spherically symmetric source.

\section{Measurement and fitting of two-particle correlations}
\begin{figure}
\centering
\includegraphics[width=0.7\textwidth]{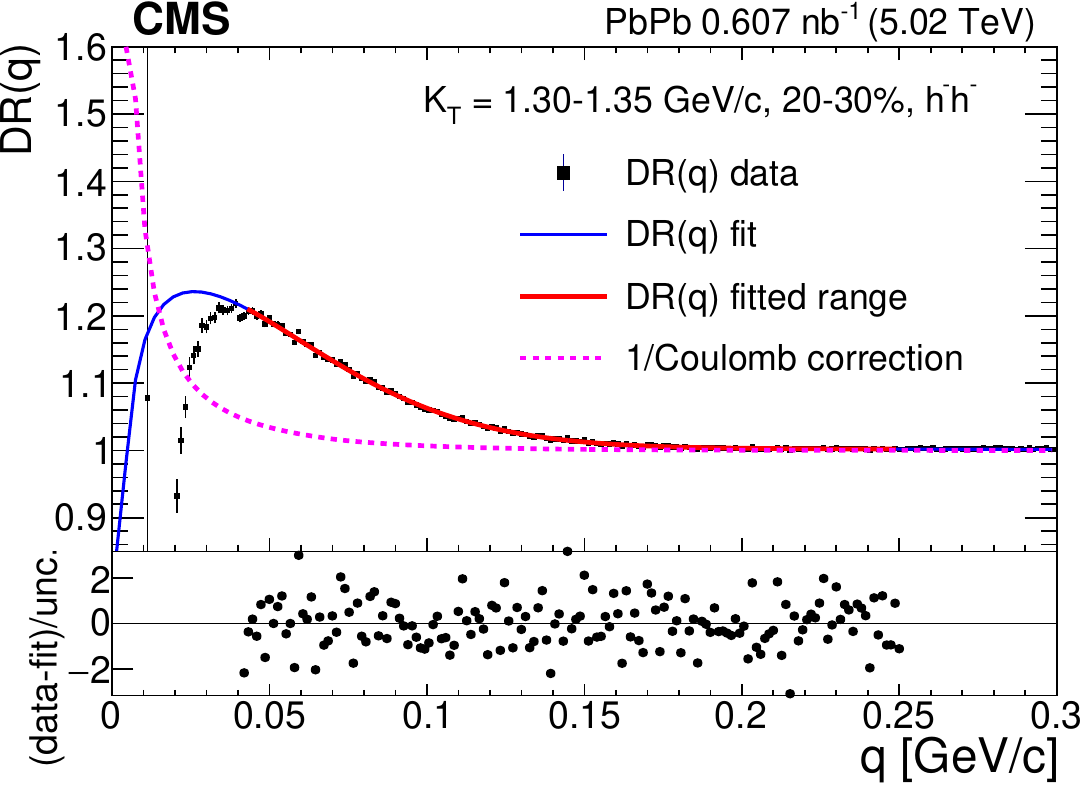}
\caption{An example fit to the correlation function $DR(q)$ in the $1.30<K_\mathrm{T}<1.35$ GeV$/c$ range, for 20--30\% centrality, reproduced from Ref.~\cite{CMS:2023xyd}. Error bars show the statistical uncertainties of the data points. The fitted function is shown in red for the fitted range, blue in the extrapolation range. The Coulomb effect is indicated in magenta.}
\label{f:DRqfit}
\end{figure}

In this analysis we investigated 4.27 billion lead-lead (PbPb) collisions at $\sqrt{s_{_\mathrm{NN}}}=5.02$ TeV, recorded by the CMS experiment in 2018. As detailed in Ref.~\cite{CMS:2023xyd}, numerous event, track and pair selection criteria were applied to ensure the purity of the investigated sample. Pair momentum distributions were divided by a background distribution, formed with a so-called mixed event sample, consisting of pairs of particles originating from different events. These remove most non-femtoscopic effects, stemming from kinematics, efficiency or phase-space region constraints. The resulting correlation functions may still contain secondary non-femtoscopic effects, such as minijets or momentum conservation. These are removed by fitting the $C(q)$ correlation function by a long-range background function and dividing by this. The result is then a double ratio, denoted as $DR(q)$. An example fit with a theoretical curve based on L\'evy-distributed source is shown in Fig.~\ref{f:DRqfit}. It is important to note that at low-$q$ values, the measurements are prone to two-track resolution effects, as proven by detector simulations~\cite{CMS:2023xyd}. Hence this part was not included in the fit, and neither was the large $q$ range, where no femtoscopic correlations are present.

Such fits have been performed in 24 ranges of pair transverse momentum ($K_\mathrm{T}$) from 0.5 to 1.9 GeV$/c$ and six centrality classes from 0 to 60\%, as detailed in Ref.~\cite{CMS:2023xyd}. The dependence of the parameters $R$, $\alpha$ and $\lambda$ on transverse mass $m_\mathrm{T}=\sqrt{m^2+K_\mathrm{T}^2}$ (assuming the pion mass for all particles, leading to a reduction in correlation strength) and centrality was investigated. Systematic uncertainties in these parameters were assessed based on variations of the event, track and pair selection criteria, as well as of the fitting range of $DR(q)$. The resulting systematic uncertainties are in the few percent range for most cases, with the largest effect coming from pair cuts and the fit range.~\cite{CMS:2023xyd} We furthermore separated the point-to-point (fluctuating) and constant part of the systematic uncertainties, with the latter yielding an overall factor for the fit parameters versus $m_\mathrm{T}$.

\section{Results and discussion}

\begin{figure}
\centering
\includegraphics[width=0.9\textwidth]{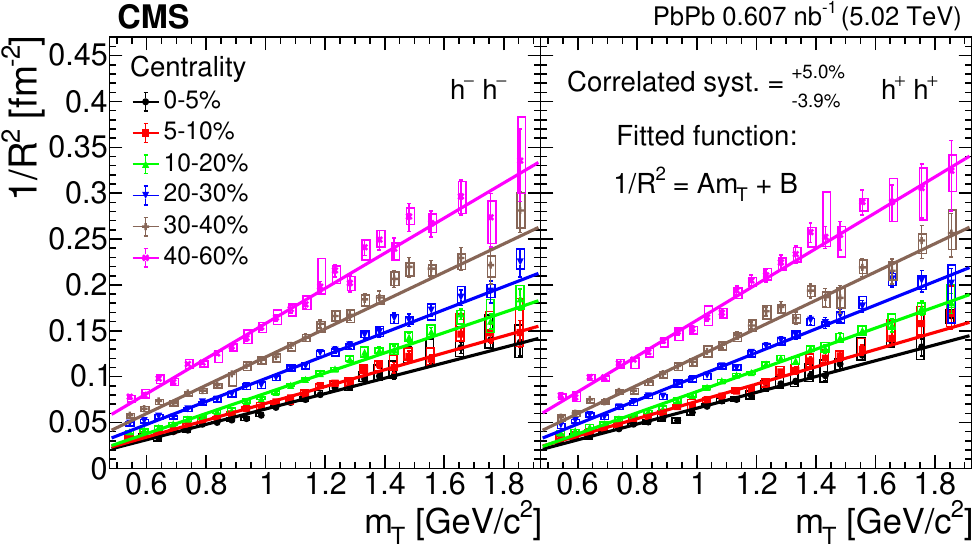}
\caption{The $1/R^2$ distribution vs transverse mass $m_\mathrm{T}$ in different centrality classes, for negatively (left) and positively (right) charged hadron pairs, reproduced from Ref.~\cite{CMS:2023xyd}. The error bars show the statistical uncertainties, while the boxes indicate the point-to-point systematic uncertainties. These boxes are slightly shifted along the horizontal axes for better visibility. The correlated systematic uncertainty is also indicated. A linear fit to the data is shown for each centrality bin.}
\label{f:1R2}
\end{figure}

The transverse mass dependence of the inverse square of the L\'evy scale $R$ is shown in Fig.~\ref{f:1R2} for all centrality classes. A linear connection between $1/R^2$ and $m_\mathrm{T}$ was predicted for Gaussian sources in Refs.~\cite{Makhlin:1987gm,Csorgo:1995bi}, hence it is interesting to observe that this connection holds for L\'evy sources as well. Further model calculations are needed in order to explain this observation. It may be furthermore noted that for more peripheral events, the $1/R^2$ values are larger, i.e., the $R$ values are smaller, in accordance with the decrease in initial system size for peripheral events.

\begin{figure}
\centering
\includegraphics[width=0.7\textwidth]{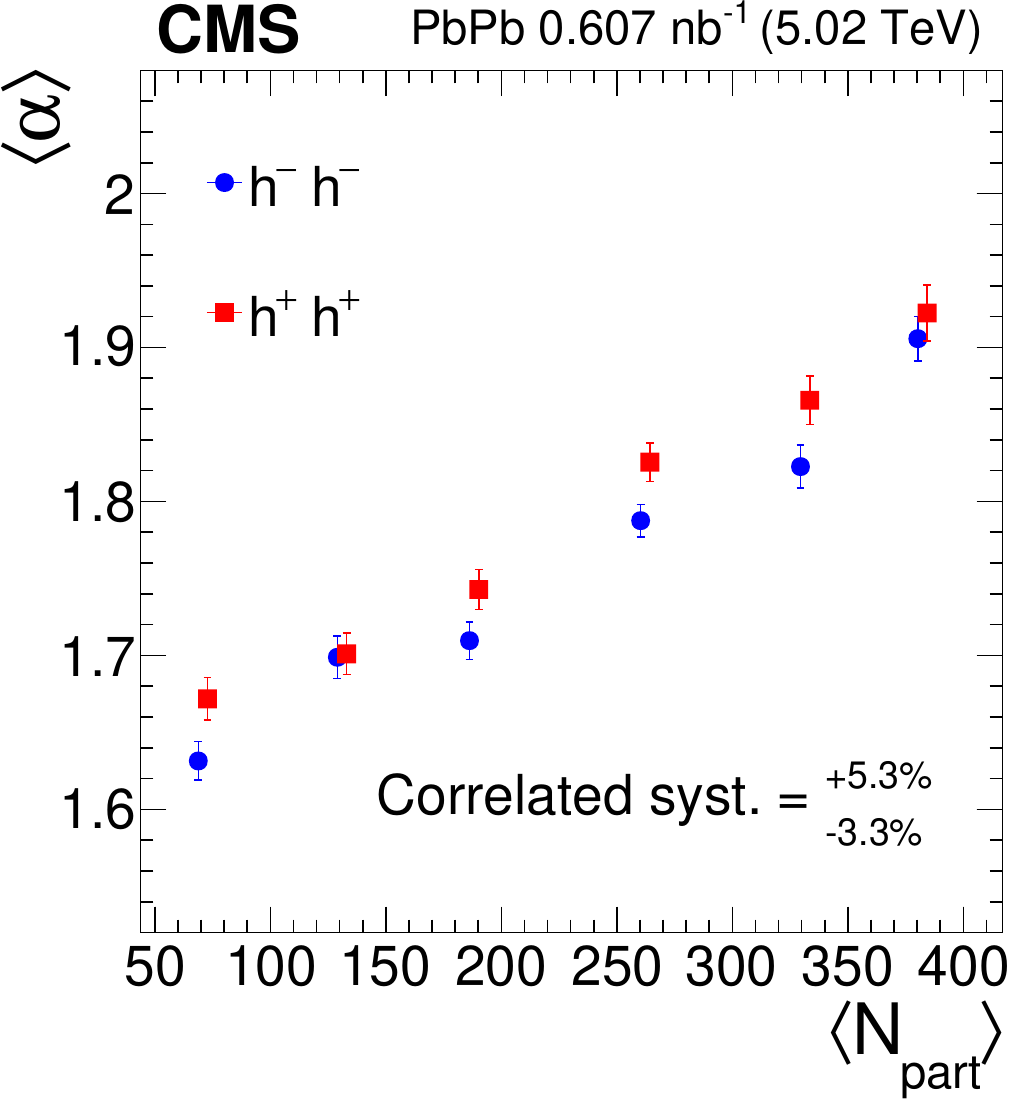}
\caption{The average L\'evy stability index $\langle\alpha\rangle$ versus $\langle N_{\rm part}\rangle$, for both positively and negatively charged hadron pairs, reproduced from Ref.~\cite{CMS:2023xyd}. The error bars show the statistical uncertainties. The correlated systematic uncertainty is also indicated. The points are slightly shifted along the horizontal axes for better visibility. }
\label{f:alphavsNpart}
\end{figure}

Figure~\ref{f:alphavsNpart} shows the centrality dependence of the L\'evy exponent ($\alpha$), averaged over $m_\mathrm{T}$ (which is meaningful to investigate due to the lack of a clear $m_\mathrm{T}$ dependence of $\alpha$, as discussed in Ref.~\cite{CMS:2023xyd}. It is apparent that for more central events, characterized by a larger $\langle N_{\rm part} \rangle$ (mean number of participants) value, the mean value of $\alpha$ is also larger. This may be connected to a larger particle density in these events, corresponding to a smaller mean free path and hence a ``less anomalous'' diffusion. Note, however, that this is in contrast to observations at RHIC.~\cite{Kincses:2024sin}

These results underline the importance of measuring the L\'evy parameters of the particle emission in ultrarelativistic nuclear collisions, and call for detailed phenomenological and theoretical investigations.

\acknowledgments
The author acknowledges the support of NKFIH grants K-138136, K-146913, and TKP2021-NKTA-64.

\end{document}